\documentclass[12pt]{article}

\usepackage{epsfig,epsf}

\newlength{\dinwidth}

\newlength{\dinmargin}

\def\lapproxeq{\lower .7ex\hbox{$\;\stackrel{\textstyle <}{\sim}\;$}}

\def\gapproxeq{\lower .7ex\hbox{$\;\stackrel{\textstyle >}{\sim}\;$}}

\def\be{\begin{equation}}

\def\ee{\end{equation}}

\def\bea{\begin{eqnarray}}

\def\eea{\end{eqnarray}}

\def\GeV{{\rm GeV}}

\begin{document}

\titlepage

\begin{flushright}

%hep-ph/0611102\\

KEK--TH--1112\\

IPPP/06/72 \\

DCPT/06/144\\

LTH 729\\

8th November 2006 \\
Revised: 20th April 2007 \\

\end{flushright}

\vspace*{1.5cm}

\begin{center}

{\Large \bf Improved predictions for $g-2$ of the muon and 
            $\alpha_{\rm QED}(M_Z^2)$}

\vspace*{1cm} {\sc K. Hagiwara}$^a$, {\sc A.D. Martin}$^b$, 
{\sc Daisuke Nomura}$^a$ and {\sc T. Teubner}$^{c}$ \\

\vspace*{0.5cm}

$^a$ {\em Theory Group, KEK, Tsukuba, Ibaraki 305-0801, Japan} \\

$^b$ {\em Department of Physics and Institute for Particle Physics Phenomenology,\\

University of Durham, Durham DH1 3LE, U.K.}\\

$^c$ {\em Department of Mathematical Sciences, University of Liverpool, Liverpool L69 3BX, U.K.}

\end{center}

\vspace*{0.5cm}

\begin{abstract}
We update the Standard Model predictions of the anomalous magnetic 
moment of the muon, $a_\mu \equiv (g-2)/2$, and the value of the 
QED coupling at the $Z$-boson mass, incorporating the new 
$e^+e^- \to \pi\pi$ data obtained by CMD-2 and KLOE, as well as 
the corrected SND data, and other improvements.  The prediction 
for $a_\mu=11659180.4 (5.1) \times 10^{-10}$ is about 
$3 \times 10^{-10}$ lower than before, and has a smaller 
uncertainty, which corresponds to a 3.4\,$\sigma$ deviation from 
the measured value. The prediction for the QED coupling is 
$\alpha(M_Z^2)^{-1}= 128.937 \pm 0.030$.
\end{abstract}

It is important to predict the anomalous magnetic moment of the muon,
$a_\mu\equiv(g_\mu-2)/2$, and the value of the QED coupling 
on the $Z$ pole as precisely as possible, in order to test the 
Standard Model and to probe New Physics. For the first quantity, we note that the Brookhaven experiment gives the average of the measurements of the $\mu^+$ and $\mu^-$ anomalous magnetic moments to be \cite{BNL}
\begin{eqnarray}
 a_\mu^{\rm exp} = 11 659 208.0(6.3) \times 10^{-10}.
 \label{eq:BNL}
\end{eqnarray}
If a statistically 
significant deviation, no matter how tiny, can be definitively 
established between the measured value $a_\mu^{\rm exp}$ and the 
Standard Model prediction, then it will herald the existence of new 
physics beyond the Standard Model. In particular the comparison offers 
valuable constraints on possible contributions from SUSY particles, see, for example, the reviews in \cite{susy}.
The second quantity, the QED coupling at the $Z$ boson mass, $M_Z$, is 
equally important. 
The uncertainty in its value is one of the major limiting factors for
precision electroweak physics. It limits, for example, the accuracy 
of the indirect estimate of the Higgs mass in the Standard Model.

The Standard Model (SM) prediction of the muon anomalous magnetic moment, $a_\mu$, may be written as the sum
of three terms,
\begin{eqnarray}
 a_\mu^{\rm SM} = a_\mu^{\rm QED} + a_\mu^{\rm EW} + a_\mu^{\rm had} .
\label{eq:amusm}
\end{eqnarray}
The QED contribution, which includes all the photonic and leptonic loops, and the EW contribution, which includes the loops involving the $W,Z$ or Higgs bosons, are known very accurately: $a_\mu^{\rm QED} = (116584718.09 \pm 0.16) \times 10^{-11}$ \cite{QED} and $a_\mu^{\rm EW} = (154 \pm 2) \times 10^{-11}$ \cite{EW}.  The main uncertainty lies in the final term, which involves the hadronic loop contributions. This term may, itself, be sub-divided into three parts
\begin{eqnarray}
 a_\mu^{\rm had} = a_\mu^{\rm had,LO}
                 + a_\mu^{\rm had,NLO}
                 + a_\mu^{{\rm had,l}\raisebox{0.45ex}{\rule{0.6ex}{0.08ex}}{\rm b}\raisebox{0.45ex}{\rule{0.6ex}{0.08ex}}{\rm l}}.
\label{eq:had}
\end{eqnarray}
At present the hadronic vacuum polarisation contributions, $a_\mu^{\rm had,LO,NLO}$, cannot be calculated sufficiently accurately from first principles \cite{AB}, but instead are evaluated using dispersion integrals over the measured cross sections for $e^+e^- \to \gamma^* \to$ hadrons. For our 2003 predictions \cite{HMNT2003} we found 
\be
a_\mu^{\rm had, LO} ~~ = ~~ (692.4 \pm 5.9_{\rm exp} \pm 2.4_{\rm rad}) ~ \times ~ 10^{-10}~, 
\label{eq:hadLO}
\ee
\be
  a_\mu^{\rm had,NLO} ~~ = ~~ (-9.79 \pm 0.09_{\rm exp} \pm 0.03_{\rm rad}) ~ \times ~ 10^{-10}~,
\label{eq:hadNLO}
\ee
where the last error corresponds to the uncertainty associated with the
radiative corrections to the cross section data. The final term in
(\ref{eq:had}), $a_\mu^{{\rm had,l}\raisebox{0.45ex}{\rule{0.6ex}{0.08ex}}{\rm
b}\raisebox{0.45ex}{\rule{0.6ex}{0.08ex}}{\rm l}}$, is the hadronic
light-by-light contribution. In our previous analysis, we took, in units
of $10^{-10}$, either\footnote{
The first is a representative value of several earlier determinations
(see, for example, the review in \cite{ny}), whereas the second value
(which was used in the note added in proof in \cite{HMNT2003}) was
obtained in \cite{MV}. In this paper we take the second value; it is
consistent with the upper bound found in \cite{es}, but see also
`Note added in proof'.}
\be
a_\mu^{{\rm had,l}\raisebox{0.45ex}{\rule{0.6ex}{0.08ex}}{\rm b}\raisebox{0.45ex}{\rule{0.6ex}{0.08ex}}{\rm l}}~~=~~8.0 \pm 4.0~~~~~~{\rm or}~~~~~~13.6 \pm 2.5.
\label{eq:lbl}
\ee

The major uncertainty in the Standard Model prediction of the
anomalous moment, $a_\mu$, comes from the contribution of the $e^+e^-
\to \pi^+ \pi^-$ channel. Indeed, the $e^+e^- \to \pi^+ \pi^-$ cross
section data, available in 2003, give the dominant contribution to
(\ref{eq:hadLO}) of about $(500 \pm 5) \times 10^{-10}$.  Since then
the situation has improved considerably. In particular, new precise
measurements have recently become available from the CMD-2 detector throughout
the centre-of-mass energy range $0.37 < \sqrt{s} < 1.38$ GeV
\cite{CMDlo,CMDmed,CMDhi}.  Moreover, the KLOE collaboration have made
measurements of the cross section by the radiative return method over
the range $0.6 < \sqrt{s} < 1$ GeV \cite{KLOE}.  Finally measurements
at the SND detector over the interval $0.4 < \sqrt{s} < 1$ GeV have
become available \cite{SND}, and subsequently have been 
corrected \cite{SNDreanal}.  Clearly all these 
data\footnote{
There is the possibility of obtaining indirect information on 
$e^+e^- \to$ hadrons in the energy range $\sqrt{s}<m_\tau$, via 
the conserved vector current (CVC) hypothesis, using the precision 
data for the hadronic decays of $\tau$ leptons. However there is 
a sizeable discrepancy between the data from the $e^+e^- \to \pi^+\pi^-$ 
experiments and that extracted from the $\pi^{\pm}\pi^0 \nu$ decay 
mode of the $\tau$. This suggests that the understanding of the CVC 
hypothesis may be inadequate at the desired level of precision,
see e.g. the recent discussion in \cite{Melnikov:2006sr}. 
Also the fair agreement between the KLOE data and the
CMD-2 and SND data strongly supports the $e^+ e^-$ 
data against the $\tau$ data~\cite{DHZ}.
In addition, $\tau$ spectral function data from Belle show a 
significant discrepancy compared to ALEPH and, to a lesser extent,
CLEO, see~\cite{hep-ex/0512071}.   
We therefore do not include the $\tau$ data in the present (or in our 
previous \cite{HMNT2003}) analysis. } 
will have an impact on the determination of $g-2$ of the muon.

Here we repeat the analysis described in detail in Ref. \cite{HMNT2003}. That is, we evaluate the dispersion relation
\be 
a_\mu^{\rm had,LO} 
 = \left(\frac{\alpha m_\mu}{3\pi}\right)^2
   \int_{s_{\rm th}}^\infty {\rm d}s\,\frac{R(s)K(s)}{s^2} \,,
\label{eq:disprel1} 
\ee
where the kernel $K(s)$ is a known function (see, for example, eq. (45) of \cite{HMNT2003}), and
\begin{equation} 
R(s) = \frac{\sigma^0_{\rm had}(s)}{\sigma_{\rm pt}(s)}\,.
\label{eq:R(s)} 
\end{equation}
The subscript 0 on $\sigma^0_{\rm had} \equiv \sigma_{\rm tot}^0(e^+e^-\to\gamma^*\to{\rm hadrons})$ is to indicate that we take the bare cross section with no initial state radiative or vacuum polarization corrections, but with final state radiative corrections; and $\sigma_{\rm pt} \equiv 4\pi\alpha^2/3s$ with $\alpha =\alpha(0)$. 

\begin{figure}
\begin{center}
\includegraphics[height=10cm]{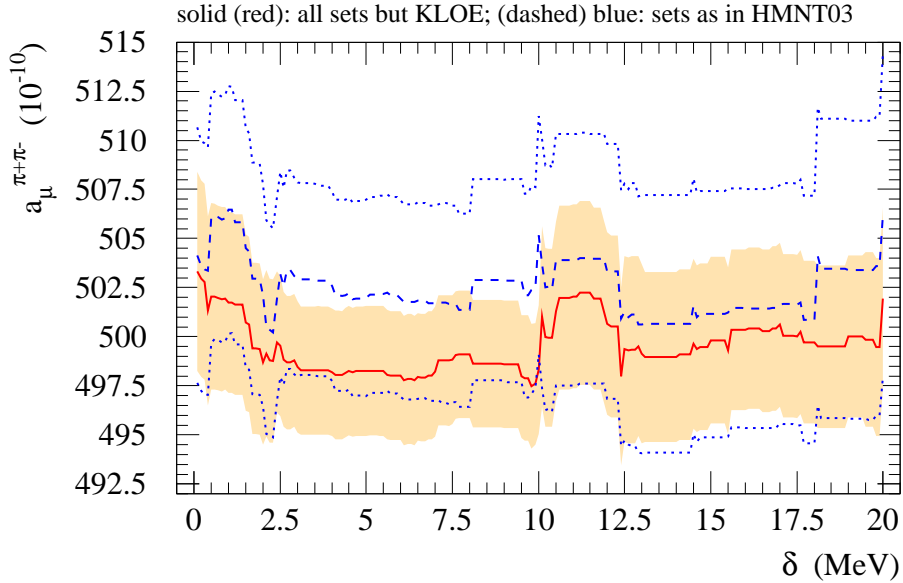}
\vspace{-1cm}
\caption{The lower continuous (red) line shows the contribution of the
  $e^+e^- \to \pi^+  \pi^-$ data in the energy region $0.32 < \sqrt{s}
  < 1.43$ GeV to $a_\mu^{\rm had,LO}$ of (\ref{eq:disprel1}), as a
  function of the data cluster size parameter $\delta$. The (orange)
  band corresponds to the uncertainty in the contribution. All the new
  $\pi^+\pi^-$ data are included except those from KLOE, see text. The upper
  dashed (blue) curve, and the error corridor given by the dotted
  lines, are the corresponding results obtained with the data as they
  were available for our previous 2003 analysis \cite{HMNT2003}. }
\label{fig:1}
\end{center}
\end{figure}
To begin, we repeat the analysis of \cite{HMNT2003} including all the
new data, except those from KLOE. The latter data have a significantly
different energy dependence, and we shall discuss their effect on the
analysis later. We combine the `bare' cross section data from the
various experiments for a given channel in clusters of a given size
$\delta$, as described in \cite{HMNT2003}. For the crucial $\pi\pi$
channel we show, in Fig.~\ref{fig:1}, how the contribution
$a_\mu^{\pi\pi,{\rm LO}}$ varies as a function of the cluster size
$\delta$. (For this channel, we have fixed the
cluster size in the small energy interval from 0.778 to 0.787 GeV
containing the $\rho-\omega$ interference effects, to be $\delta=1$
MeV, thus improving the quality and stability of the fit
considerably.) The improvement due to
the inclusion of the new $\pi\pi$ data is immediately clear from
Fig.~\ref{fig:1}. Choosing the cluster size $\delta$ to be 3.5 MeV, as
in our previous analysis \cite{HMNT2003}, we see that the contribution
$a_\mu^{\pi\pi,{\rm LO}}$ is reduced by $4.6 \times 10^{-10}$, and that
the uncertainty in the value decreases from $5.0 \times 10^{-10}$ to
$3.2 \times 10^{-10}$. Moreover the result is stable to the variation
of the choice of the cluster size in the range from 3 to 10
MeV. Furthermore the minimum $\chi^2$ of the clustering fit to the
data, divided by the number of degrees of freedom, $\chi^2_{\rm
  min}/{\rm d.o.f.}$, is, to a good approximation, equal to $1.1$
for all choices of $\delta$ in this interval.
\begin{figure}
\begin{center}
\includegraphics[height=11cm]{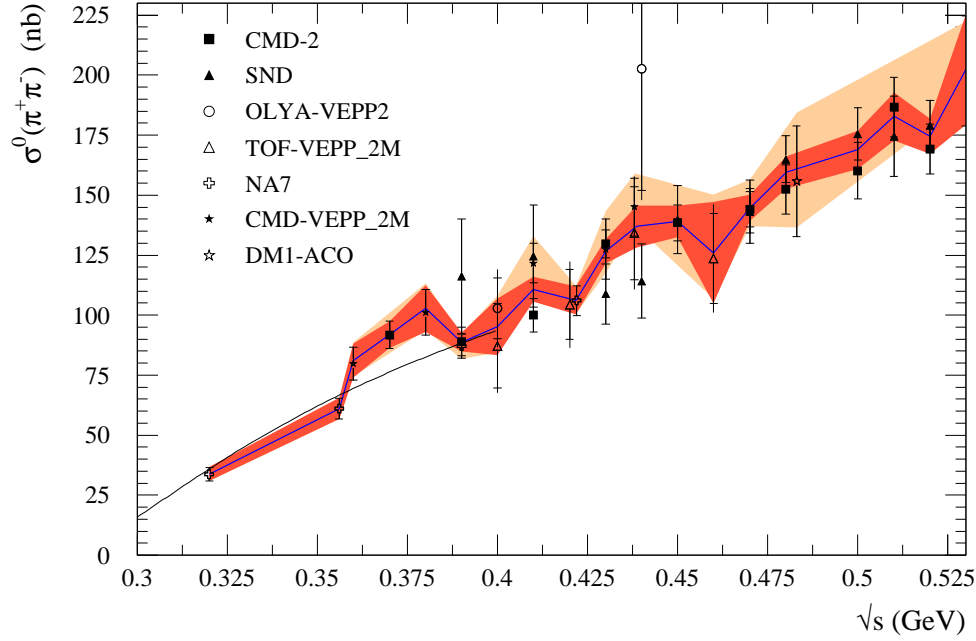}\\
\vspace{-2cm}
\includegraphics[height=11cm]{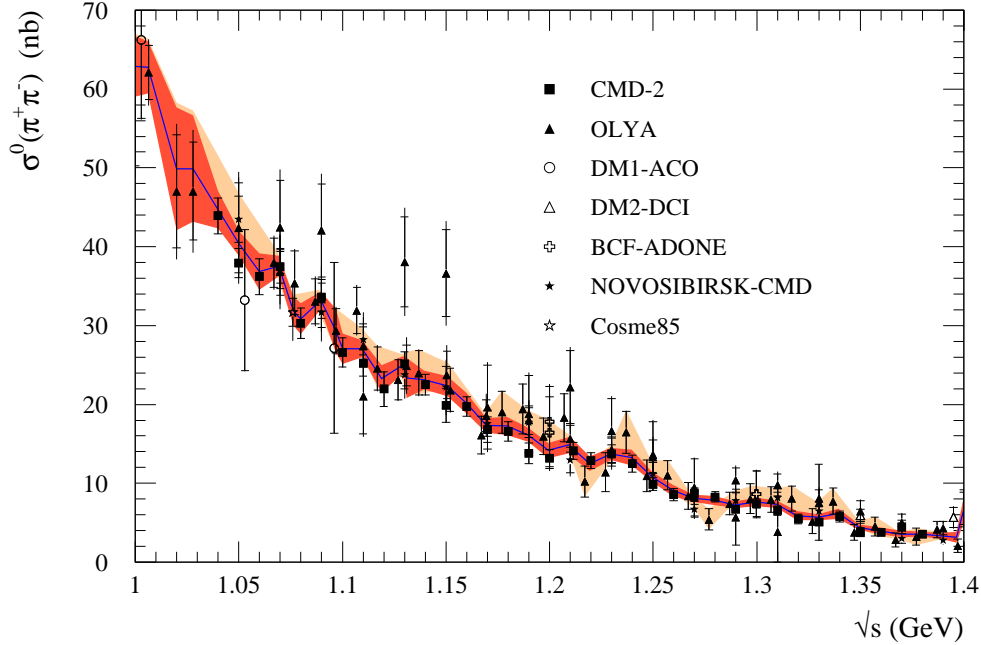}
\vspace{-1cm}
\caption{$\pi\pi$ data in the low and high energy $\rho$-resonance
  tail regions (upper and lower panels), compared to our fits using a
  clustering size of $\delta=4.0$ MeV. The light (orange) band
  indicates the error band of the fit without the recent CMD-2 (2006)
  and SND data which became available after our previous analysis
  \cite{HMNT2003}, whereas the (blue) line and darker (red) band
  correspond to the mean value and error band of the new fit including
  all data as indicated in the two panels.}
\label{fig:2}
\end{center}
\end{figure}
It is informative to trace the origin of the reduction of $4.6 \times
10^{-10}$ in $a_\mu$ due to the addition of the new $\pi\pi$ data. It
comes about equally from the intervals $0.32 < \sqrt{s} < 0.6$ GeV and
$1 < \sqrt{s} < 1.43$ GeV, with only a small reduction in $a_\mu$
coming from the intervening energy range. This becomes clear from
Fig.~\ref{fig:2}, where the data are displayed together with two fits
for a cluster size $\delta=4.0$ MeV (recall from Fig.~\ref{fig:1} the
insensitivity of the integrated result to variation of $\delta$ in the
range $3-10$ MeV), which correspond to our previous and the present
analyses. A comparison of the two bands shows that the introduction of
the CMD-2 (and SND) data gives, on average, a smaller cross section in
the energy regions above 410 and above 1030 MeV in the upper and lower plots, respectively, of Fig.~\ref{fig:2}.

We now discuss the inclusion of the KLOE $\pi\pi$ data in the
analysis. Note that KLOE measures the hadronic cross section via the
{\em radiative return} method in $e^+ e^- \to \phi \to \pi\pi\gamma$
at the $\Phi$ factory DA$\Phi$NE in Frascati.  Here the observation of
initial state photon radiation at various energies allows for a
determination of the invariant mass spectrum of the $\pi\pi$ system.
This analysis is completely independent of the `direct scan'
measurements of CMD-2 and SND at VEPP-2M in Novosibirsk which use a
tunable $e^+e^-$ beam energy. Unfortunately, the KLOE data \cite{KLOE}
have a different energy dependence to the other $\pi\pi$ data sets,
especially when compared with the recent CMD-2 and SND analyses, see
e.g. the discussions in \cite{SNDreanal,Sibidanov:2006ts}. Our
clustering prescription allows overlapping data sets to adjust by an
overall constant within the systematic error of each set, but does not
allow for an energy dependent renormalization\footnote{
We have also studied the effect of an energy dependent renormalization,
using either a linear or a constant plus logarithmic form;
neither of these forms were able to improve the fit with the KLOE
data significantly.}. 
If we would include the KLOE
data in the fit, then they would be normalized upwards by nearly two
percent.\footnote{This is because there are fewer data points in the
  low energy region lying high (compared to CMD-2) than lying low in
  the central peak and high energy tail region, and because KLOE's
  quoted systematic error is about twice as large as CMD-2's.}
\begin{figure} [ht]
\begin{center}
\includegraphics[height=14cm]{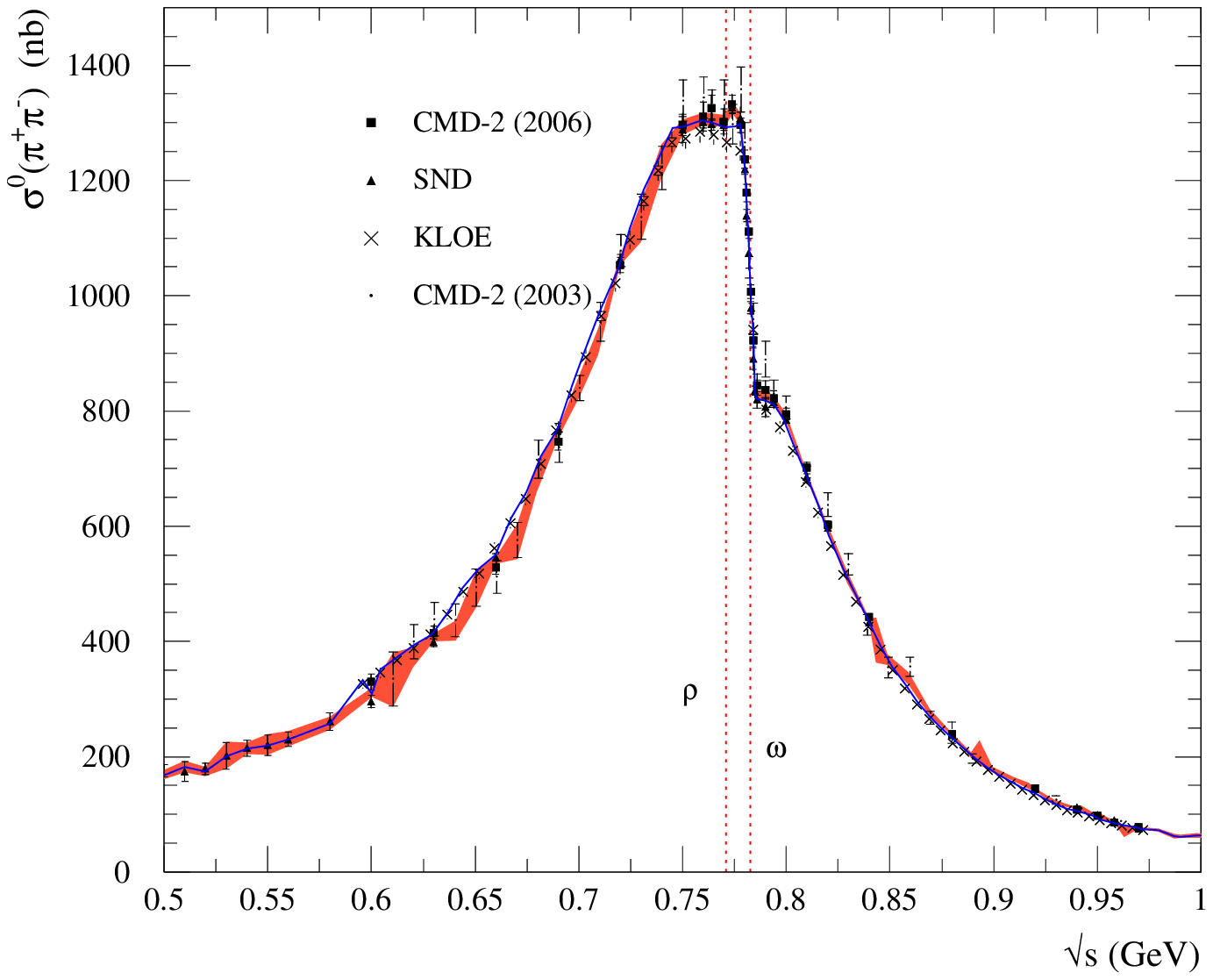}
\vspace{-1cm}
\caption{Mean value of the fit of all $\pi\pi$ data in the $\rho$ central 
region (blue line) compared to the fit without the KLOE data (red band), 
together with data from CMD-2, SND and KLOE (as indicated).}
\label{fig:3}
\end{center}
\end{figure}
This in turn would lead to a sizeable effect in the fit, see
Fig. \ref{fig:3}: the (blue) line is the mean value of the fit including the
KLOE data, whereas the (red) band is the error band for the fit excluding
KLOE. In the dispersion integral this would lead to an artificial overall shift
of $a_{\mu}^{\pi\pi}$ upwards by about $5\times 10^{-10}$. As the
$\chi^2_{\rm min}/{\rm d.o.f.}$, for the `clustering' fit to the $\pi\pi$ data, would increase significantly from the value 1.1 found above, the improvement of
the error on $a_{\mu}^{\pi\pi}$ after inflation by
$\sqrt{\chi^2_{\rm min}/{\rm d.o.f.}}$ would be very small.  However the KLOE
data \cite{KLOE} more than adequately populate the energy interval of
their measurement, $0.60 < \sqrt{s} < 0.97$ GeV, so that we can
compute reliably their integrated contribution to the dispersion
relation of (\ref{eq:disprel1}). We find 
\be
a_\mu^{\pi\pi,\rm LO}({\rm KLOE\ only}, 0.6-0.97 ~\GeV) 
~~=~~(385.7 \pm 4.9) \times 10^{-10}.
\ee
%which is in agreement with the number quoted in \cite{KLOE}.
This should be compared to that obtained in the complete analysis (in which the KLOE data were omitted)
\be
a_\mu^{\pi\pi,\rm LO}({\rm fit\ w/o\ KLOE}, 0.6-0.97 ~\GeV) 
~~=~~(384.3 \pm 2.5) \times 10^{-10},
\ee
where, as is to be expected, the error is less.  Moreover, despite
their different energy dependence, the agreement of the KLOE
integrated contribution with that of the fit of all other data is
excellent. Therefore, for our SM prediction of $a_{\mu}$, we combine
both contributions for this energy interval. Doing so we arrive at
\be
a_\mu^{\pi\pi,\rm LO} ~~=~~(384.6 \pm 2.2) \times 10^{-10},
\ee
for the interval $0.60 < \sqrt{s} < 0.97$ GeV. The effect of the 
KLOE data is to slightly increase the magnitude and reduce the error.  As a result, 
the $\pi\pi$ contribution in the interval $0.32 < \sqrt{s} < 1.43$ GeV, 
becomes $(498.5 \pm 2.9) \times 10^{-10}$. 

In addition to the inclusion of the new $\pi\pi$ data, we have made other improvements in the analysis. One small improvement that we have made is to evaluate the vacuum polarisation corrections to the data using our recent determination \cite{HMNTalpha} of the running of the effective coupling $\alpha(q^2)$ in the time-like region $q^2=s$, which was obtained using our clustered data set. This is more consistent than using the evaluation by Jegerlehner \cite{Jegerlehner} that we did previously.  This is a correction to a correction, and so hardly changes the result. 

A more important improvement is that for the subleading exclusive channels we have included new data from CMD-2
\cite{cmd2newsubl} ($2\pi^+2\pi^-$, $\pi^0\gamma$, $\eta\gamma$),
BaBar \cite{babarnewsubl} ($\pi^+\pi^-\pi^0$, $2\pi^+2\pi^-$,
$3\pi^+3\pi^-$, $\pi^+\pi^- K^+ K^-$, $p\bar p$); and from BES
\cite{besnewincl} for the inclusive hadronic cross section.  The effects of the new data are
summarized in Table~\ref{tab:gm2table}. It lists the contributions to $a_\mu^{\rm had, LO}$ which have changed since our previous prediction. As expected, the main change arises from the inclusion of the new $\pi\pi$ data: $-4.32$ in units of $10^{-10}$. This is partially compensated by changes in other contributions, such as $+0.84$ (due mainly to the new BES data) and $+0.60$ (due to the new leptonic widths of $J/\psi$ and $\psi'$ \cite{RPP06}).  
Including all the hadronic contributions we obtain
\be
a_\mu^{\rm had, LO} ~~ = ~~ 
(689.4 \pm 4.2_{\rm exp} \pm 1.8_{\rm rad}) ~ \times ~ 10^{-10}~, 
\label{eq:hadLOnew}
\ee
and, similarly for NLO, 
\be
  a_\mu^{\rm had,NLO} ~~ = ~~ 
(-9.79 \pm 0.08_{\rm exp} \pm 0.03_{\rm rad}) ~ \times ~ 10^{-10}~,
\label{eq:hadNLOnew}
\ee
which are to be compared with (\ref{eq:hadLO}) and (\ref{eq:hadNLO}) 
respectively.

\begin{table}[ht]\begin{center}
\begin{tabular}{l|l|l|l}
channel & range (GeV) & $a_{\mu} \times 10^{10}$ & HMNT03 \cite{HMNT2003} \\ \hline
$\pi^+\pi^-$     &$0.32 - 1.43$ & $498.46 \pm 2.87$  &$502.78 \pm 5.02$ \\
$\pi^+\pi^-\pi^0$&$0.66 - 1.43$ &  $46.18 \pm 0.94$ & $46.43 \pm 0.90$ \\
$2\pi^+2\pi^-$   &$0.6125 - 1.43$&  $6.01 \pm 0.19$ & $6.16 \pm 0.32$ \\
$\pi^0\gamma$    &$0.60 - 1.03$ &   $4.54 \pm 0.12$ & $4.50 \pm 0.15$ \\
$\eta\gamma$     &$0.69 - 1.35$ &   $0.72 \pm 0.03$ & $0.73 \pm 0.03$ \\
inclusive        &$1.43 - 11.09$ &  $74.80 \pm 2.67$ & $73.96 \pm 2.68$ \\
$J/\psi+\psi'$   &               &  $7.90 \pm 0.16$ & $7.30 \pm 0.43$ \\
\end{tabular}
%\vspace{2mm} 
\caption{Contribution of different channels to $a_{\mu}$ compared to
  the numbers as given in \cite{HMNT2003}. }
\label{tab:gm2table} 
\end{center}
\end{table}

\begin{figure}[ht]
\begin{center}
\includegraphics[height=11cm]{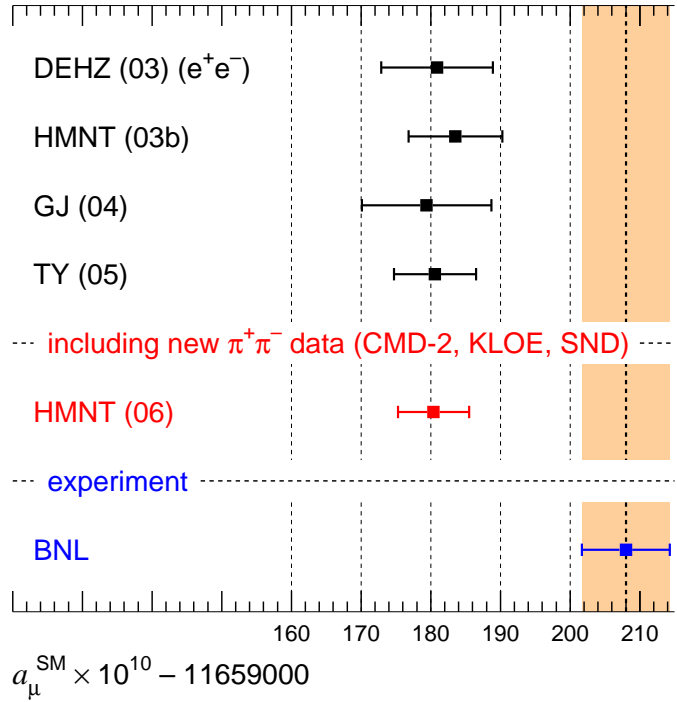}
\vspace*{-1.cm}
\caption{In the lower half of the plot we compare the experimental value
 \cite{BNL} of the anomalous magnetic moment of the muon, $(g-2)/2
 \equiv a_\mu$, with the SM prediction calculated in the text. We see
 that there is now a 3.4\,$\sigma$ discrepancy between experiment and the SM prediction. In the upper part of the plot we show four earlier determinations \cite{HMNT2003, early} of $a_\mu$ that were made before the CMD-2 \cite{CMDlo,CMDmed,CMDhi}, KLOE \cite{KLOE} and SND \cite{SND,SNDreanal} $e^+e^- \to \pi\pi$ data became available. The light-by-light contribution used in these determinations varies from $(8.6 \pm 3.5) \times 10^{-10}$ in DEHZ(03) to the more recent estimate \cite{MV} of $(13.6 \pm 2.5) \times 10^{-10}$ used in HMNT(03b,06), see (\ref{eq:lbl}).}
\label{fig:4}
\end{center}
\end{figure}
Finally, adding all the terms of (\ref{eq:amusm}) and 
(\ref{eq:had}), we obtain the updated Standard Model prediction for 
the anomalous magnetic moment of the muon:
\begin{equation}
     a_\mu^{\rm SM} = (11 659 180.4 \pm 5.1) \times 10^{-10}.
\label{eq:a_mu_totalSM}
\end{equation}
That is, the difference $\delta a_\mu (\equiv 
a_{\mu}^{\rm exp} - a_{\mu}^{\rm SM})$
is  $\delta a_\mu = (27.6 \pm 8.1) \times 10^{-10}$.
Eq.~(\ref{eq:a_mu_totalSM}) should be compared to our previous prediction (HMNT(03b)), given in the note added in proof in \cite{HMNT2003}, of
\begin{equation}
     a_\mu^{\rm SM} = (11 659 183.5 \pm 6.7) \times 10^{-10}.
\end{equation}
The new data, particularly for $e^+e^- \to \pi\pi$ 
in the centre-of-mass energy range $0.37 < \sqrt{s} < 1.38$ GeV, 
have decreased the value of $a_\mu^{\rm SM}$ by $3.1 \times 10^{-10}$ 
and reduced the error from $6.7 \times 10^{-10}$ to 
$5.1 \times 10^{-10}$.  Both of these effects increase the disagreement 
with the measured value. The situation is shown pictorially in 
Fig.~\ref{fig:4}. We now have a discrepancy of 3.4\,$\sigma$, which is 
larger than before\footnote{
If we do not use the KLOE data at all, then we have
\begin{equation}
   a_{\mu}^{\rm had,LO} = 
          (689.2 \pm 4.3_{\rm exp} \pm 1.8_{\rm exp}) \times 10^{-10},
\end{equation}
in place of (\ref{eq:hadLOnew}).  (The change in the hadronic NLO 
contribution is invisible to this accuracy.)
The corresponding value for the total SM prediction is
\begin{equation}
    a_{\mu}^{\rm SM} = (11659180.2 \pm 5.3) \times 10^{-10}.
\end{equation}
The deviation $\delta a_\mu$ from the experimental value 
$a_\mu^{\rm exp}$ is
\begin{equation}
    \delta a_{\mu} = a_\mu^{\rm exp} - a_\mu^{\rm SM}
        = (27.8 \pm 8.2) \times 10^{-10},
\end{equation}
which again corresponds to a 3.4\,$\sigma$ discrepancy.}$^{,}$\footnote{
A preliminary DEHZ analysis \cite{ichep}, 
also including the new data, finds 
$a_\mu^{\rm SM} = (11 659 180.5 \pm 5.6) \times 10^{-10}$, for which they quote a 3.3\,$\sigma$ discrepancy.}. Moreover, note that we use the 
recent, larger, value \cite{MV} of the light-by-light contribution, 
near the upper limit estimated in \cite{es}, which reduces the
discrepancy.  

The larger discrepancy, $a_\mu^{\rm exp}-a_\mu^{\rm SM}$, 
which arises from the inclusion of the new $\pi\pi$ data, is becoming 
a more significant indication of New Physics beyond the Standard Model.  
The effect of supersymmetry on $a_\mu$ can be seen, for example, from 
the reviews in \cite{susy}. Finally, it is worth noting that the 
theoretical error is now below the experimental error on $a_\mu$. 
With further measurements of the low energy hadronic cross sections 
underway, which will improve the accuracy of $a_\mu^{\rm SM}$, the 
case for improving the measurement of $a_\mu$ is strong.

\begin{figure}[ht] \begin{center}
\includegraphics[height=8cm]{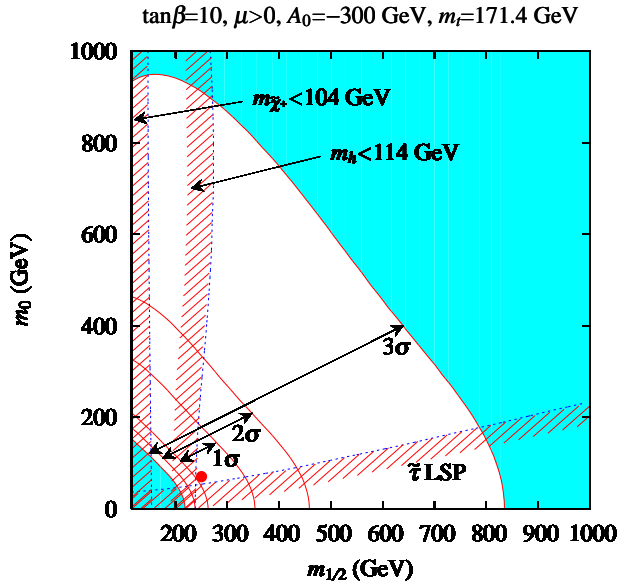}
\caption{The contours in the minimal SUGRA 
$m_{1/2}$, $m_0$ plane allowed by the difference 
$a_{\mu}^{\rm exp} - a_{\mu}^{\rm SM} = (27.6 \pm 8.1) 
\times 10^{-10}$ at 1-$\sigma$, 2-$\sigma$ and 3-$\sigma$. 
The other parameters are $\tan\beta=10$, $A_0=-300$ GeV,
$\mu>0$, and the SPS 1a$^\prime$ reference point is shown by the big dot.
Also shown are the constraints from the Higgs 
boson and the chargino masses from the direct searches 
and the charged dark matter ($\tilde{\tau}$ LSP).}
\label{fig:mSUGRAplot}
\end{center} \end{figure}

As an example of the impact of the difference 
$a_\mu^{\rm exp}-a_\mu^{\rm SM}$, we show in Fig.~\ref{fig:mSUGRAplot} 
the allowed region of the minimal SUGRA parameters $m_{1/2}$ and 
$m_0$.  We choose the other parameters to be $\tan\beta=10$, 
$A_0=-300$ GeV, and $\mu>0$, so that the commonly studied reference 
point SPS 1a$^\prime$~\cite{SPA} appears as the big dot on the plane.
We also plotted the constraints arising from the Higgs boson mass and 
the chargino mass from the negative results of direct searches 
at LEP~\cite{RPP06}, and that from the lightest charged SUSY 
particle (the region labeled by ``$\tilde{\tau}$ LSP'').  
(To make this figure, we used the program ``{\tt SuSpect}''~\cite{SuSpect}
to calculate the mass spectrum of the SUSY particles except
for the lightest Higgs boson mass, for which we used
``{\tt FeynHiggs}''~\cite{FeynHiggs}.)

The new data also improve the predicted value of the QED coupling 
at the $Z$ boson mass. The value of $\alpha(M_Z^2)$ is obtained 
from \cite{PDG2006}
\be
\alpha^{-1}\equiv \alpha(0)^{-1} = 137.035999710(96)
\label{eq:alpha^{-1}}
\ee
using the relation
\be 
\alpha(M_Z^2)^{-1} = 
 \left( 1 - \Delta\alpha_{\rm lep}(M_Z^2)
          - \Delta\alpha_{\rm had}(M_Z^2)    
 \right) \alpha^{-1}, 
\ee
where $\Delta\alpha_{\rm lep}(M_Z^2)=0.03149769$, and $\Delta\alpha_{\rm had}$ is evaluated from the dispersion relation
\begin{equation} 
  \Delta \alpha_{\rm had}(M_Z^2) 
 = -\frac{M_Z^2}{4\pi^2\alpha} 
  {\rm P} \int_{s_{\rm th}}^\infty {\rm d} s\,
  \frac{\sigma_{\rm had}^0(s)}{s-M_Z^2}\,. 
\label{eq:disprel2} 
\end{equation}
The {\it bare} cross section $\sigma_{\rm had}^0(s)$ is defined below (\ref{eq:R(s)}). In comparison with (\ref{eq:disprel1}), dispersion relation (\ref{eq:disprel2}) is less sensitive to the values of $\sigma_{\rm had}^0(s)$ in the low energy region. We therefore do not expect the error on $\alpha(M_Z^2)$ to reduce quite as much as that on $a_\mu^{\rm had,LO}$, when we use the improved $e^+e^- \to \pi\pi$ data.

It is conventional to separate out the top-quark contribution and to write
\be
\Delta\alpha_{\rm had}~=~\Delta\alpha_{\rm had}^{(5)}+\Delta\alpha^{\rm top},
\ee
where, using $m_t=171.4 \pm 2.1$ GeV, perturbative QCD determines 
$\Delta\alpha^{\rm top}(M_Z^2) = -0.000073 (02)$.  
We proceed as in Ref.~\cite{HMNT2003}. Using the new clustered data 
to evaluate the dispersion relation (\ref{eq:disprel2}), we find 
\begin{eqnarray}
 \Delta\alpha_{\rm had}^{(5)}(M_Z^2) 
&=& 0.02768\pm 0.00017_{\rm exp}\pm 0.00013_{\rm rad} \\
&=& 0.02768\pm 0.00022. 
\end{eqnarray}
This, in turn, gives
\be 
\alpha(M_Z^2)^{-1} = 128.937 \pm 0.030, 
\ee
which should be compared to our previous estimate of 
$\alpha(M_Z^2)^{-1} = 128.954 \pm 0.031.$ The accuracy 
is now $23\times 10^{-5}$, which is still the least accurately 
determined of the three 
fundamental parameters of the electroweak theory;
$\Delta G_\mu/G_\mu= 0.9\times 10^{-5}$  and 
$\Delta M_Z/M_Z= 2.3\times 10^{-5}$, where $G_\mu$ is the Fermi constant.

{\it Note added in proof:} After submission of this letter a new assessment
of the status of the light-by-light contributions has appeared
\cite{Bijnens}. If we would use their estimate of
$a_{\mu}^{\rm had,l\mbox{-}b\mbox{-}l}=(11 \pm 4)\times 10^{-10}$, 
the difference
$a_{\mu}^{\rm exp} - a_{\mu}^{\rm SM}$ would slightly widen to $(30.2 \pm
8.7)\times 10^{-10}$, corresponding to a 3.5\,$\sigma$ discrepancy.

\section*{Acknowledgements}

We would like to thank Achim Denig and Simon Eidelman for valuable discussions
concerning the hadronic data.  TT thanks the UK Particle Physics and
Astronomy Research Council for an Advanced Fellowship.

\end{document}